\documentclass[preprint,prd]{revtex4} 
 
\begin{document} 
\input{epsf}

\title{Frequency Spectra and Probability Distributions for Quantum
Fluctuations}

\author{ L.H. Ford} 
 \email[Email: ]{ford@cosmos.phy.tufts.edu} 
 \affiliation{Institute of Cosmology  \\
Department of Physics and Astronomy\\ 
         Tufts University, Medford, MA 02155}

\begin{abstract} 
This paper discusses two distinct, but related issues in quantum 
fluctuation effects. The first is the frequency spectrum which can be 
assigned to one loop quantum processes. The formal spectrum is a flat
one, but the finite quantum effects can be associated with a rapidly 
oscillating spectrum, as in the case of the Casimir effect. The leads
to the speculation that one might be able to dramatically change the 
final answer by upsetting the delicate cancellation which usually
occurs. The second issue is the probability distribution for quantum
fluctuations. It is well known that quantities which are linear in a free
quantum field have a Gaussian distribution. Here it will be argued that
quadratic quantities, such as the quantum stress tensor, must have a
skewed distribution. Some possible implications of this result for
inflationary cosmology will be discussed. In particular, this might be
a source for non-Gaussianity.
\end{abstract}

\maketitle 
 
\baselineskip=14pt 

\section{Introduction}

This paper will deal with two aspects of quantum fluctuations. One will
be the frequency spectrum which is associated with Casimir energy,
that is, with the expectation value of the stress tensor operator
or other quadratic operators. Some earlier work will be reviewed,
in which it was shown that the frequency spectra are wildly oscillating 
functions, which nonetheless have finite integrals. The finite Casimir
energy  always corresponds to a very small fraction of the area under 
one oscillation peak. The possibility of altering this remarkable 
cancellation will be discussed.

The second aspect of quantum fluctuations to be considered is the
probability distribution associated with the fluctuations of smeared
operators. The well-known result of a Gaussian distribution for 
linear operators will be rederived. It will then be shown by way
of a simple example that the fluctuations of quadratic operators
can be described by a skewed, and hence non-Gaussian, distribution.
Some further comments and speculations on the generic form of the
probability distribution for stress tensor fluctuations will be offered.

\section{Frequency Spectra}

It is well-known that the formal frequency spectrum of quantum fluctuations
is flat; all modes appear on an equal basis in the expansion of a quantum
field operator. This leads to the formally divergent zero point energy
\begin{equation}
E_0 = \sum_\lambda \frac{1}{2}\, \hbar\, \omega \,. \label{eq:zero_point}
\end{equation}
However, this ``unprocessed'' spectrum is unobservable. The only quantities
which can be observed arise from ``processed'' spectra, which have been 
modified by a physical process. The formation of a black hole, for example,
processes the flat spectrum of incoming vacuum fluctuations and converts it
into a Planck spectrum of outgoing particles, the Hawking radiation.
The finite effects of one loop quantum processes may also be associated 
with a nontrivial frequency spectrum. Consider the Casimir effect as an 
example. The presence of boundaries modifies
the divergent vacuum energy, Eq.~(\ref{eq:zero_point}), by a finite amount,
the Casimir energy. For a scalar field which satisfies periodic boundary 
conditions in one of three spatial dimensions, this energy is
(Units in which $\hbar = c = 1$ will be used in the remainder of this 
paper.)
\begin{equation}
E_C = -\frac{\pi^2\, A}{90\, L^3} \, ,  \label{eq:Casimir_energy}
\end{equation}
where $L$ is the periodicity length and $A$ is the transverse area. 

It is of interest to ask whether one can assign a finite frequency
spectrum to the Casimir energy, that is, find a function $\sigma(\omega)$
such that 
\begin{equation}
E_C = \int_0^\infty d\omega\, \sigma(\omega) \,.
\end{equation}
This was done in Refs.~\cite{Ford88,Hacyan}, 
where it was shown that one can obtain $\sigma(\omega)$
as a Fourier transform of the renormalized energy density operator at 
time-separated points. A more recent discussion is given in Ref.~\cite{Lang}.
Let $T_{\mu\nu}(t-t')$ be the renormalized stress 
tensor operator evaluated at two points separated in time by $t-t'$. Then
$T_{\mu\nu}(0)$, the operator evaluated at coincident points, is the
observable stress-energy associated with the Casimir effect. In particular,
\begin{equation}
E_C =  A\, L\, T_{tt}(0) \,.
\end{equation} 
We can write 
\begin{equation}
\sigma(\omega) = \frac{A\, L}{2 \pi} \, \int_{-\infty}^{\infty} 
{\rm e}^{i \omega t}\, dt \,.
\end{equation}
This function was evaluated explicitly for the electromagnetic case in
Ref.~\cite{Hacyan}, and for the case of a scalar field in Ref.~\cite{Ford88}. 
In the latter case, the result is
\begin{equation}
\sigma(\omega) = -\frac{A\, \omega^2}{\pi\, L} \, S(\omega L) \, ,
                                      \label{eq:sigma_C}
\end{equation}
where 
\begin{equation}
S(x) = \sum_{n=1}^\infty \frac{\sin(n x)}{n}
\end{equation}
is a discontinuous, periodic function given by $S(x+2\pi)=S(x)$
and
\begin{equation}
S(x) = \frac{1}{2}(\pi - x)\, , \qquad 0 \leq x < 2\pi \,.
\end{equation}

\begin{figure}
\begin{center}
\leavevmode\epsfysize=6cm\epsffile{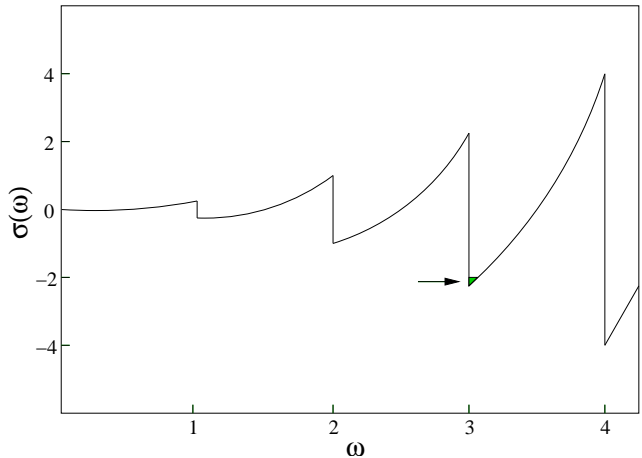}
\end{center}
\caption{ The frequency spectrum, $\sigma(\omega)$, given by 
Eq.~(\ref{eq:sigma_C}) for the Casimir
energy.  The oscillations almost exactly cancel, leaving
a net area under the curve equal to that of the shaded region indicated
by the arrow. Here $\sigma$ is in units of $2A/L$ and $\omega$ in units
of $2\pi/L$. } 
\label{fig:Casimir}
\end{figure}

At first sight, the integral of $\sigma(\omega)$ over all frequencies is 
poorly defined. However, it can be defined with a suitable convergence
factor, such as an exponential function. One finds that the Casimir energy 
arises in the limit in which the convergence factor is removed. For example,
\begin{equation}
E_C = \lim_{\beta \to 0} \int_0^\infty d\omega\, 
{\rm e}^{-\beta \omega} \, \sigma(\omega) \,.
\end{equation}
The result is that the contributions of different frequency intervals
almost exactly cancel one another, leaving a finite result which is small
compared to the area of each of the peaks in Fig.~\ref{fig:Casimir}.

\begin{figure}
\begin{center}
\leavevmode\epsfysize=6cm\epsffile{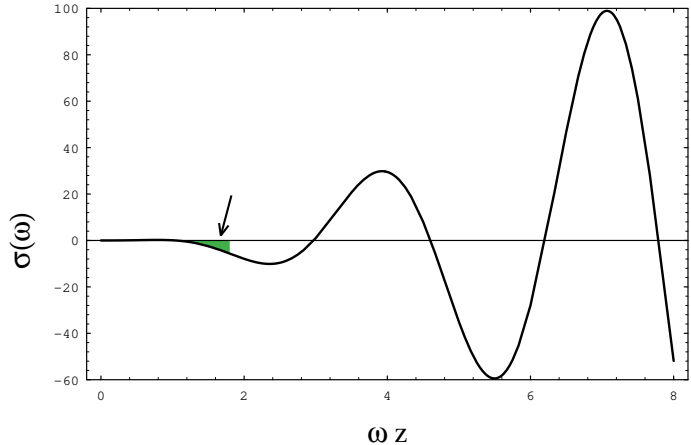}
\end{center}
\caption{The frequency spectrum, $\sigma(\omega)$, for the Casimir-Polder
potential.  The oscillations again almost exactly cancel, and
the net area is equal to that of the shaded region indicated
by the arrow.} 
\label{fig:CP}
\end{figure}

A similar spectrum can be associated with the asymptotic Casimir-Polder 
potential between a polarizable particle, such as an atom in its ground
state, and a perfectly reflecting wall. In the limit that the atom is far
from the plate, compared to the wavelength associated with the transition 
between the ground state and first excited state, Casimir and 
Polder~\cite{CP} showed that the interaction energy is, in Gaussian units,
\begin{equation}
V_{CP} = - \frac{3\, \alpha_0}{8 \pi\, z^4}\,,   \label{eq:CP}
\end{equation}
where $z$ is the distance to the wall, and $\alpha_0$ is the static 
polarizability of the atom. It may be expressed as
\begin{equation}
V_{CP} = \frac{\alpha_0}{4 \pi \, z^3}\, 
\int_0^\infty d \omega\, \sigma(\omega) \,,
\end{equation}
where, in this case,
\begin{equation}
\sigma(\omega) =  \Bigl[(2\,\omega^2\,z^2 -1) \sin 2\omega z
+ 2\,\omega\,z \cos 2\omega z \Bigr] \,.
\end{equation}
As before, the integral on $\omega$ may be defined using a convergence factor.
The frequency spectrum in this case is illustrated in Fig.~\ref{fig:CP}.

As we have seen, the frequency spectra associated with Casimir-type
effects can be wildly oscillatory, and yet tend to integrate to a
small net energy. An obvious question is whether it may be possible
to slightly modify the contribution of specific frequency intervals
and upset the delicate cancellation between the different positive 
and negative peaks in Figs.~\ref{fig:Casimir} and \ref{fig:CP}.
For example, if one could alter the reflectivity of parallel plates
just in a selected frequency range, it would seem to be possible
to have a Casimir force much larger than that between perfectly
reflecting plates, and which could be repulsive as well as 
attractive~\cite{Ford93}. A large, repulsive Casimir force could have
significant industrial applications, such as allowing nearly frictionless
bearings. Unfortunately, it is not at all straightforward to enhance
Casimir forces. One might replace perfectly reflecting plates by
dielectric slabs, which will have a finite, frequency dependent reflectivity.
However, the Lifshitz theory~\cite{Lifshitz} predicts 
that the force between two
dielectric half-spaces will always be attractive and smaller in magnitude
that the force in the perfectly reflecting limit. At least part of the 
reason for this is that the dielectric function $\varepsilon(\omega)$ 
must satisfy the
Kramers-Kronig relations, which follow from the requirement of analyticity
in the upper-half $\omega$ plane. This analyticity property is in turn
a consequence of causality~\cite{Jackson}. Analyticity constrains
$\varepsilon(\omega)$ in such a way that the contributions of various
frequency regions continue to cancel efficiently. If one were to allow
non-analytic dielectric functions, then it becomes possible to 
construct situations in which the Lifshitz thoery predicts 
repulsive forces~\cite{Ford93}.

Another approach to attempt to enhance Casimir forces was taken in 
Ref.~\cite{Ford98}. There the force due to the quantized electromagnetic field
on a small dielectric sphere near a perfectly reflecting plate was calculated
A quasi-oscillatory result was obtained, which can be either attractive
or repulsive, and is much larger in magnitude than the Casimir-Polder
force,  from Eq.~(\ref{eq:CP}), for a non-dispersive sphere. In 
Ref.~\cite{VF04}, it was shown that the qualitative form of the 
result does not 
depend upon whether the plate is perfectly reflecting or not. In both
cases. the sphere was taken to have a plasma model dielectric function, 
and the resulting force oscillations as a function of distance from the plate
have a scale set by the associated plasma wavelength. One can think of
the effect of the sphere's nontrivial frequency response as  upsetting the 
cancellations that occur for a  non-dispersive sphere, and which are 
illustrated in Fig.~\ref{fig:CP}.

If the vacuum modes of the electromagnetic field gave the only contribution
to the force on the sphere, then there would be a large repulsive force on 
the sphere at certain separations which would be large enough to levitate 
the sphere in the earth's gravitational field, and should be observable.  
However, there is another contribution to the net force coming from 
quantum mechanical fluctuations of the electric charge within the sphere.
This is the effect of what is sometimes called the plasmonic modes. It 
was suggested by Barton~\cite{Barton05} that the effect of the plasmonic modes
should cancel the quasi-oscillatory terms coming from the vacuum modes.
This suggestion is in fact correct, and was recently confirmed by an
explicit calculation~\cite{Ford05}. There is still a nonzero net force, but
it is always attractive and of the order of magnitude of that for a
non-dispersive sphere. The cancellation occurs only if the sphere plasmon
modes are in their ground state. If they are excited, then the 
quasi-oscillatory behavior appears in the net force. This is similar
to the case of an atom. If the atom is in its ground state, then the
force is attractive at all separations~\cite{CP}. If
the atom is in an excited state, however, the net force is 
quasi-oscillatory and typically much larger in magnitude than in the
ground state~\cite{Barton74}. It is still unclear why the plasmon modes
can cancel the effects of the vacuum modes so efficiently, and whether one 
can engineer materials with repulsive Casimir forces without exciting
plasma oscillations.

\section{Probability Distributions}

In this section, we will turn to a somewhat different measure of vacuum
fluctuations, the probability distribution associated with the fluctuations.
First consider the case of a quantity which is linear in a free quantum field.
It is well known that such quantities have a Gaussian distribution of 
probabilities. However, it will be worthwhile illustrating this result with
an explicit example. Let $\varphi(x)$ be a free scalar field and let
\begin{equation}
\bar{\varphi} = \int \varphi(x)\, dV
\end{equation}
be a smeared field operator averaged over some spacetime region. The dimension
of the spacetime is not relevant here. We could also include a sampling 
function $f(x)$ and write $f(x)\,dV$ in place of $dV$ without changing the 
result. Next consider the expectation value
of a power of $\bar{\varphi}$ in the vacuum state. All of the odd powers
have vanishing expectation value because the $n$-point functions vanish for
all odd $n$. That is,
\begin{equation}
\langle \bar{\varphi}^{2n+1} \rangle = \int \langle \varphi_1
\varphi_2\cdots \varphi_{2n+1} \rangle \, dV_1 dV_2 \cdots dV_{2n+1} =0\,,
\end{equation}
where $\varphi_1 = \varphi(x_1)$, ect. However, all of the even moments
are nonzero and can be expressed as powers of the second moment,
\begin{equation}
\langle \bar{\varphi}^{2} \rangle = \int \langle \varphi_1
\varphi_{2} \rangle \,.
\end{equation}
For example, the identity
\begin{equation}
\langle \varphi_1 \varphi_2 \varphi_3 \varphi_4 \rangle =
\langle \varphi_1 \varphi_2 \rangle \langle \varphi_3 \varphi_4 \rangle
+ \langle \varphi_1 \varphi_3 \rangle \langle \varphi_2 \varphi_4 \rangle
+ \langle \varphi_1 \varphi_4 \rangle \langle \varphi_2 \varphi_3 \rangle
     \label{eq:4point}
\end{equation}
leads to the result
\begin{equation}
\langle \bar{\varphi}^4 \rangle\ = 3 \langle \bar{\varphi}^{2} \rangle^2 \,.
\end{equation}
This is a special case of the general result
\begin{equation}
\langle \bar{\varphi}^{2n} \rangle\ = (2n-1)!! \, 
\langle \bar{\varphi}^{2} \rangle^n \,. \label{eq:moment2n}
\end{equation}
The latter result can be obtained from the following counting argument.
Wick's theorem allows us to decompose the $2n$-point function 
$\langle \varphi_1 \varphi_2 \cdots \varphi_{2n} \rangle$ into a sum of
products of two-point functions, just as is illustrated in 
Eq.~(\ref{eq:4point}). All that we need to know is the number of terms in
this sum.  The first contraction of the $2n$-point function contain $2n-1$
terms, as we can select any field to start, and it then has $2n-1$ partners
with which it can be contracted. Similarly, the contraction of each 
$(2n-2)$-point function to $(2n-4)$-point functions contains $2n-3$ terms,
and so on, leading to the 
factor of $(2n-1)!!$ in Eq.~(\ref{eq:moment2n}).

This result for the general even moment implies that the probability 
distribution is a Gaussian function
\begin{equation}
P(\bar{\varphi}) = \frac{1}{\sqrt{2 \pi\, \langle \bar{\varphi}^{2} \rangle}}\;
{\rm e}^{-2 \bar{\varphi}^2/\langle \bar{\varphi}^{2} \rangle} \,,
\end{equation}
which is confirmed by the facts that
\begin{equation}
\langle \bar{\varphi}^{2} \rangle = \int_{-\infty}^\infty x^2 \,P(x)\, dx
\end{equation}
and that
\begin{equation}
\langle \bar{\varphi}^{2n} \rangle = \int_{-\infty}^\infty x^{2n} \,P(x)\, dx
= (2n-1)!! \, \langle \bar{\varphi}^{2} \rangle^{n} \,.
\end{equation}
The result that $P(\bar{\varphi})$ is a Gaussian function is responsible 
for the
prediction of Gaussian fluctuations in inflationary cosmology. In most versions
of inflation, density perturbations are linked to the quantum fluctuations of 
a scalar inflaton field, which is treated as a free field~\cite{Peebles}. 
This leads to
a Gaussian distribution of density fluctuations, which seems to be consistent
with observation~\cite{CNSTZ}.

Let us now turn to the fluctuations of quadratic operators, such as the
smeared stress tensor. Now there is no particular reason to expect the
probability distribution to be symmetric, much less Gaussian. In fact, 
as we will see from an explicit example, the distribution is in general
a skewed one. The example will involve a massive scalar field in 
two-dimensional Minkowski spacetime, but in the limit of a very small mass.
The Hadamard function for this field is
\begin{equation}
G(x,x') = \frac{1}{2}\,\{\varphi(x), \varphi(x')\} = 
- \frac{1}{4}\, N_0(m \sqrt{\Delta t^2 - \Delta x^2}) \, ,
\end{equation} 
where $m$ is the mass and $N_0$ is a Neumann function. In the limit of
small $m$, this function becomes
\begin{equation}
G(x,x') = - \frac{1}{4 \pi} \, \ln[ \mu^2 (\Delta t^2 - \Delta x^2)]\,, 
\end{equation}
where $\mu = {\rm e}^\gamma\, m/2$ and $\gamma$ is Euler's constant.
The commutator function, in the limit that $m \rightarrow 0$, is
\begin{equation}
G_C(x,x') = [\varphi(x), \varphi(x')] 
= \frac{i}{4}\, \left[ \theta(\Delta x - \Delta t) 
- \theta(\Delta x - \Delta t) \right]\, ,
\end{equation}
where $\theta(x) = 1$ for $x >0$ and $\theta(x) = -1$ for $x <0$.
Note that the commutator function is finite for $m =0$, whereas the
Hadamard function has a logarithmic divergence in this limit. This is a 
well-known feature of the massless scalar field in two-dimensional
Minkowski spacetime. This divergence can be removed by selecting a vacuum
state which breaks Lorentz invariance~\cite{FV86}, but here we will assume
a small, nonzero mass. 

Next consider the average of the normal ordered square of $\varphi$ over 
a finite time interval
\begin{equation}
\bar{\varphi^2} = \frac{1}{T}\, \int_0^T :\varphi(t)^2:\, dt \,.
\end{equation} 
In this model, all of the field operators will be taken to be at the 
same point in space, so that $\Delta x = 0$. We wish to calculate the vacuum
expectation value of various powers of $\bar{\varphi^2}$. Begin with
the second moment,
\begin{equation}
\langle (\bar{\varphi^2})^2 \rangle = \frac{1}{T^2}\,
 \int_0^T \, dt_1  \int_0^T \, dt_2 
\langle:\varphi(t_1)^2:\, :\varphi(t_2)^2:\rangle  
=  \frac{2}{T^2}\,
 \int_0^T \, dt_1  \int_0^T \, dt_2 
\langle \varphi(t_1)\, \varphi(t_2)\rangle^2 \, ,
\end{equation}
where the last step follows from Wick's theorem. The unsymmetrized
two-point function can be written as a sum of the Hadamard and commutator
functions
\begin{equation}
\langle \varphi(t_1)\, \varphi(t_2)\rangle = 
G(t_1,t_2) + \frac{1}{2}\, G_C(t_1,t_2)\,. 
\end{equation}
We can now evaluate the second moment explicitly and find 
\begin{equation}
\langle (\bar{\varphi^2})^2 \rangle = \frac{1}{2 \pi^2}\, \ln^2(T \mu)  
\end{equation}
in the limit that $T \mu \ll 1$. In this limit, the Hadamard function
part of the two-point function gives the leading contribution.

Next we turn to the third moment
\begin{equation}
\langle (\bar{\varphi^2})^3 \rangle = \frac{1}{T^3}\,
 \int_0^T \, dt_1  \int_0^T \, dt_2 \int_0^T \, dt_3 \,
\langle:\varphi(t_1)^2:\, :\varphi(t_2)^2: :\varphi(t_3)^2: \rangle \, , 
\end{equation}
which can be written as
\begin{equation}
\langle (\bar{\varphi^2})^3 \rangle = \frac{8}{T^3}\,
\int_0^T \, dt_1  \int_0^T \, dt_2 \int_0^T \, dt_3 \,
\langle \varphi(t_1)\, \varphi(t_2)\rangle 
\langle \varphi(t_2)\, \varphi(t_3)\rangle
\langle \varphi(t_1)\, \varphi(t_3)\rangle \,.
\end{equation}
In the limit of small $T \mu$, this becomes
\begin{equation}
\langle (\bar{\varphi^2})^3 \rangle = -\frac{1}{\pi^3}\, \ln^3(T\mu)
= \left[\frac{1}{\pi}\, \ln\left( \frac{1}{T \mu}\right) \right]^3 \,.
\end{equation}
The third moment is positive $\langle (\bar{\varphi^2})^3 \rangle > 0$,
and of the same order of magnitude as the second moment in the sense
that
\begin{equation}
\langle (\bar{\varphi^2})^3 \rangle^\frac{1}{3} = \sqrt{2}\,
\langle (\bar{\varphi^2})^2 \rangle^\frac{1}{2} \,.
\end{equation}
Thus the probability distribution is significantly skewed.

\begin{figure}
\begin{center}
\leavevmode\epsfysize=6cm\epsffile{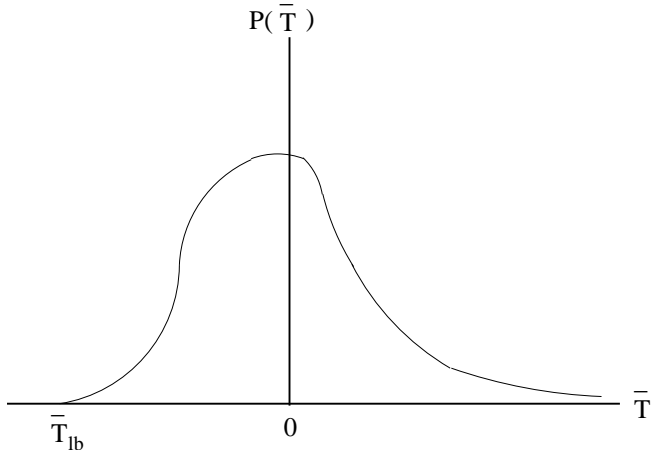}
\end{center}
\caption{A possible probability distribution $P(\bar{T})$ for stress
tensor fluctuations is illustrated as a function of an averaged stress
tensor component $\bar{T}$. This distribution is zero  below the lower bound, 
$\bar{T} < \bar{T}_{\rm lb} < 0$, but has a nonzero tail extending infinitely
far in the positive direction. The majority of the area under the curve lies
in the negative region, $\bar{T} < 0$.} 
\label{fig:sketch}
\end{figure}

The corresponding calculations for components of the quantum stress tensor
have not yet been performed, but it is reasonable to guess that the results
will be similar to those of this $\varphi^2$ model in two dimensions.
The technical details are somewhat more complicated, however. Let
\begin{equation}
\bar{T} = \int :T_{tt}: \,f(x)\, dV
\end{equation}
be the average of the normal-ordered energy density over a spacetime region
defined by the non-negative sampling function $f(x)$. 
The stress tensor correlation
function is singular in the limit of null separated points,
\begin{equation}
\langle :T_{tt}(x): \,:T_{tt}(x'): \rangle \sim (x-x')^{-2d} \, ,
\end{equation}
where $d$ is the number of spacetime dimensions. Thus the integral for the 
second moment $\langle \bar{T}^2 \rangle$ will appear to contain a
non-integrable singularity. However, it is possible to define such
integrals using an integration by parts procedure~\cite{WF01,FW03,FR05}. 
It is also possible
to use dimensional regularization~\cite{FW04}. 
In both approaches, the integral becomes 
finite. The third and higher moments can be defined in a similar manner.
If $\langle \bar{T}^3 \rangle > 0$, as in the case of the $\varphi^2$ mode,
then the probability distribution will again be skewed. There should be
a connection between this distribution and the quantum inequalities
which set lower bounds on the expectation value of $\bar{T}$ in non-vacuum
states~\cite{F78,F91,FR95,FR97,FLAN,PF971,PFGQI,FE,Fewster} . 
In the case of temporal averaging only, these inequalities are of the form
\begin{equation}
\langle \bar{T} \rangle \geq \bar{T}_{\rm lb} = - \frac{C}{\tau^d} \, ,
\end{equation}
where $\tau$ is the characteristic width of the sampling function
and $C$ is a constant. 
Thus negative energy densities are allowed in quantum field theory, but 
are tightly constrained. The lower bound in the expectation value in an
arbitrary quantum state, and the lower bound on the probability
distribution of fluctuations in the vacuum state are expected 
to coincide~\cite{FFR}. 
One way to understand this is to imagine writing the Minkowski vacuum state
as a superposition of an alternative basis set of states. In each of these
states, the measured $\bar{T}$ is bounded below by $\bar{T}_{\rm lb}$.
Hence the possible fluctuations which can be observed in the 
Minkowski vacuum state are also bounded below by the same value. 
If the probability distribution $P(\bar{T})$ has positive skewness and
a finite negative lower bound at $\bar{T} = \bar{T}_{\rm lb}$, then
we can make some general observations on $P(\bar{T})$, The first moment is 
defined to vanish, $\langle \bar{T} \rangle = 0$. A positive third moment
implies a long tail in the positive direction. Thus the greater portion 
of the area will lie to the left of $\bar{T} = 0$. Thus a typical measurement 
of the sampled energy density is more likely to yield a negative than a
positive value. However, when the measured value is positive, its magnitude
is likely to be greater than when it is negative. Figure~\ref{fig:sketch} 
is a sketch of a hypothetical probability distribution which satisfies 
all of these conditions. A more detailed analysis of the probability
distribution for various cases is currently in progress~\cite{FFR}.
 
The non-Gaussian nature of quantum stress tensor fluctuations may have
applications to inflationary cosmology in the form of a source of
a non-Gaussian component in the density perturbations. This is currently
under investigation~\cite{WNF}.

\begin{acknowledgments}
I have benefitted from discussion with several colleagues, including
G. Barton, C.J. Fewster, T.R. Roman and V. Sopova.
  This work was supported in part by the National
Science Foundation under Grant PHY-0244898.
\end{acknowledgments}

\end{document}